\documentclass[lettersize,journal]{IEEEtran}
\usepackage{amsmath,amsfonts}
\usepackage{algorithmic}
\usepackage{algorithm}
\usepackage{array}
\usepackage[caption=false,font=normalsize,labelfont=sf,textfont=sf]{subfig}
\usepackage{textcomp}
\usepackage{stfloats}
\usepackage{url}
\usepackage{verbatim}
\usepackage{graphicx}
\usepackage{subcaption}
\usepackage{caption}
\usepackage{subcaption}
\usepackage{cite}
\usepackage{tikz}
\begin{document}

\title{Near-Optimal Low-Complexity MIMO Detection via
Structured Reduced-Search Enumeration}

\author{Logeshwaran Vijayan
\thanks{The algorithm behind this paper was originally invented/implemented, applied for patent and granted in 2005, 2007 and 2011 respectively while the author was with Redpine Signals Inc., which was later acquired by Silicon Labs. However, this paper does an independent analysis of it.}
}



\maketitle

\begin{abstract}
Maximum-likelihood (ML) detection in high-order MIMO systems is computationally prohibitive due to exponential complexity in the number of transmit layers and constellation size. In this white paper, we demonstrate that for practical MIMO dimensions (up to $8\times8$) and modulation orders, near-ML hard-decision performance can be achieved using a structured reduced-search strategy with complexity linear in constellation size. Extensive simulations over i.i.d. Rayleigh fading channels show that list sizes of $3 |\mathcal{X}|$ for $3\times3$, $4 |\mathcal{X}|$ for $4\times4$, and $8 |\mathcal{X}|$ for $8\times8$ systems closely match full ML performance, even under high channel condition numbers, $|\mathcal{X}|$ being the constellation size. In addition, we provide a trellis based interpretation of the method. We further discuss implications for soft LLR generation and FEC interaction.
\end{abstract}


\section{Introduction}
\IEEEPARstart{M}{IMO} detection remains a fundamental challenge in modern wireless systems due to the exponential growth of the ML search space: $ |\mathcal{X}|^{N_t} $, where $N_t$ is the number of transmit layers and $|\mathcal{X}|$ is the constellation size.

While sphere decoding and lattice-reduction techniques reduce average complexity, worst-case behavior remains prohibitive for hardware implementation. This motivates structured reduced-search detectors that exploit channel geometry, ordering, and diversity properties.

This work builds on that motivation and demonstrates that \emph{exact ML hard decisions} can often be recovered using dramatically smaller candidate lists.

\section{System Model}
We consider a narrowband flat-fading MIMO system:
\begin{equation}
\mathbf{y} = \mathbf{H}\mathbf{x} + \mathbf{n},
\end{equation}
where $\mathbf{H} \in \mathbb{C}^{N_r \times N_t}$ is the channel matrix, $\mathbf{x} \in \mathcal{X}^{N_t}$ is the transmit vector and $\mathbf{n} \sim \mathcal{CN}(0, \sigma^2 \mathbf{I})$.

Applying QR decomposition (with possible column permutation $\mathbf{P}$),
\begin{equation}
\mathbf{H}\mathbf{P} = \mathbf{Q}\mathbf{R},
\end{equation}
where $\mathbf{Q}$ is unitary and $\mathbf{R}$ is upper triangular.
Left-multiplying by $\mathbf{Q}^H$ gives
\begin{equation}
\tilde{\mathbf{y}} = \mathbf{R}\tilde{\mathbf{x}} + \tilde{\mathbf{n}},
\end{equation}
with
\[
\tilde{\mathbf{y}} = \mathbf{Q}^H \mathbf{y}, \quad
\tilde{\mathbf{x}} = \mathbf{P}^T \mathbf{x}, \quad
\tilde{\mathbf{n}} = \mathbf{Q}^H \mathbf{n}.
\]

Expanding above equation,
\begin{equation}
\tilde{y}_i = \sum_{j=i}^{N_t} r_{ij} x_j + \tilde{n}_i,
\quad i = 1, \dots, N_t,
\end{equation}
which is directly analogous to a causal ISI channel with memory $N_t - i$.

Maximum-likelihood detection therefore corresponds to minimizing the global metric
\begin{equation}
\hat{\mathbf{x}} =
\arg\min_{\mathbf{x} \in \mathcal{S}^{N_t}}
\sum_{i=1}^{N_t}
\left|
\tilde{y}_i - \sum_{j=i}^{N_t} r_{ij} x_j
\right|^2,
\end{equation}

\section{Proposed Detector}

The main idea behind this paper was applied for patent in 2007 and granted in 2011 [1]. To the best of author's knowledge, there is a similar/closely related method that was proposed around the same time-frame by another set of authors [2]. In this paper, we provide a different interpretation/explanation of the method. In that sense, this work represents an independent theoretical analysis. The results are based on computer simulations of publicly disclosed algorithmic principles [1], [2] and do not represent the proprietary implementation of any specific commercial product. To accurately reflect its operational characteristics, we refer to this detector as the \emph{Multi-Pivot Multiple-Hypothesis Trellis-Like MIMO Detector (MP-MHT-MD)}.

The detector interprets the QR-transformed MIMO system as a spatial ISI channel and performs
a depth-wise hypothesis search analogous to a trellis decoder. Unlike conventional sphere
decoding, which prunes paths early based on partial metrics from a single pivot layer, the
proposed method cyclically enumerates each spatial layer as a pivot and preserves multiple
competing hypotheses until sufficient decoding depth is reached. This delayed-pruning strategy
ensures that near-ML paths are not prematurely eliminated, resulting in ML-equivalent hard
decisions with drastically reduced search complexity.

We explain the algorithm here for $3 \times 3$ MIMO system. 
\begin{align}
y_3 &= r_{33} x_3 + n_3 \\
y_2 &= r_{22} x_2 + r_{23} x_3 + n_2 \\
y_1 &= r_{11} x_1 + r_{12} x_2 + r_{13} x_3 + n_1
\end{align}
\IEEEpubidadjcol
We pivot on $x_3$ and search over all possible hypothesis. For each hypothesis of $x_3$, out of the $|\mathcal{X}|$ possible $x_2$ candidates, the best $x_2$ minimizing equation (8) is selected and then the same operation is repeated for last layer $x_1$ where we find best $x_1$ for each of the $|\mathcal{X}|$ surviving hypothesis of $(x_3,x_2)$ pairs.

\subsection{Spatial Trellis Interpretation}
ML detection corresponds to sequence estimation over a spatial ISI channel. Each layer represents a decoding stage, and nodes represent constellation hypotheses. The ML path emerges only after full trellis depth is explored; early pruning may eliminate the correct path.


Equations (7) to (9) are similar to a 3-tap ISI channel as given below 
\begin{equation}
\begin{aligned}
y_0 &= h_0 x_0 \\
y_1 &= h_0 x_1 + h_1 x_0 \\
y_2 &= h_0 x_2 + h_1 x_1 + h_2 x_0
\end{aligned}
\end{equation}

and hence reveal a \emph{causal spatial ISI structure}:
\begin{itemize}
\item Symbol $x_3$ interferes with observations of $x_2$ and $x_1$
\item Symbol $x_2$ interferes with observations of $x_1$
\item Only $x_1$ is interference-free at full depth
\end{itemize}

Thus, MIMO detection after QR decomposition is equivalent to sequence estimation
over a spatial ISI channel, where the ``time'' axis corresponds to the decoding order.

\subsubsection{Implication for Detection Algorithms}

The spatial ISI interpretation implies that:
\begin{itemize}
\item ML detection is a \emph{sequence estimation} problem, not independent symbol slicing
\item Early pruning (e.g., sphere decoding) may eliminate the true ML path prematurely
\item Retaining competing hypotheses until sufficient depth is essential
\item Multiple decoding orders correspond to multiple spatial trellises
\end{itemize}

This perspective naturally motivates reduced-complexity, list-based,
and multi-order detection strategies that preserve ML hard decisions
while significantly reducing computational complexity.

Define branch metrics:
\begin{align}
\mathrm{BM}_3(x_3) &= |y_3 - r_{33} x_3|^2 \\
\mathrm{BM}_2(x_2,x_3) &= |y_2 - r_{22} x_2 - r_{23} x_3|^2 \\
\mathrm{BM}_1(x_1,x_2,x_3) &= |y_1 - r_{11} x_1 - r_{12} x_2 - r_{13} x_3|^2
\end{align}

Path metric at depth k: 
\begin{align}
\mathrm{PM}_k = \sum_{i=3}^{k} \mathrm{BM}_i 
\end{align}

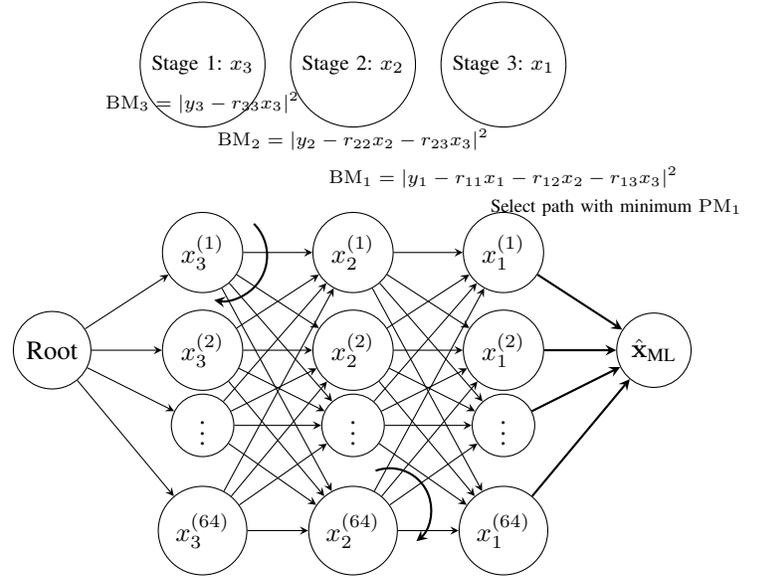
\begin{figure}
\begin{tikzpicture}[
    node distance=0.5cm,
    every node/.style={circle, draw, minimum size=3mm},
    metric/.style={rectangle, draw=none, font=\scriptsize},
    >=stealth
]

\node (root) {Root};

\node[right of=root, xshift=1.5cm,yshift=1.3cm] (x3a) {$x_3^{(1)}$};
\node[right of=root, xshift=1.5cm]              (x3b) {$x_3^{(2)}$};
\node[right of=root, xshift=1.5cm,yshift=-1.0cm] (x3c) {$\vdots$};
\node[right of=root, xshift=1.5cm,yshift=-2.4cm] (x3d) {$x_3^{(64)}$};

\node[right of=x3b, xshift=1.5cm,yshift=1.3cm] (x2a) {$x_2^{(1)}$};
\node[right of=x3b, xshift=1.5cm]              (x2b) {$x_2^{(2)}$};
\node[right of=x3b, xshift=1.5cm,yshift=-1.0cm] (x2c) {$\vdots$};
\node[right of=x3b, xshift=1.5cm,yshift=-2.4cm] (x2d) {$x_2^{(64)}$};

\node[right of=x2b, xshift=1.5cm,yshift=1.3cm] (x1a) {$x_1^{(1)}$};
\node[right of=x2b,xshift=1.5cm]              (x1b) {$x_1^{(2)}$};
\node[right of=x2b, xshift=1.5cm,yshift=-1.0cm] (x1c) {$\vdots$};
\node[right of=x2b, xshift=1.5cm,yshift=-2.4cm] (x1d) {$x_1^{(64)}$};

\node[right of=x1b, xshift=1.5cm] (ml) {$\hat{\mathbf{x}}_{\text{ML}}$};

\foreach \x in {x3a,x3b,x3c,x3d}
  \draw[->] (root) -- (\x);

\foreach \x in {x3a,x3b,x3c,x3d}
{
  \draw[->] (\x) -- (x2a);
  \draw[->] (\x) -- (x2b);
  \draw[->] (\x) -- (x2c);
  \draw[->] (\x) -- (x2d);
}

\foreach \x in {x2a,x2b,x2c,x2d}
{
  \draw[->] (\x) -- (x1a);
  \draw[->] (\x) -- (x1b);
  \draw[->] (\x) -- (x1c);
  \draw[->] (\x) -- (x1d);
}

\draw[->, thick]
  ([xshift=3mm]x3a.north east)  
  arc[start angle=45, end angle=-100, radius=6mm];

\draw[->, thick]
  ([xshift=3mm,yshift=2mm]x2d.north)  
  arc[start angle=110, end angle=-50, radius=5.5mm];

\foreach \x in {x1a,x1b,x1c,x1d}
  \draw[->, thick] (\x) -- (ml);

\node[metric, above of=x3a, yshift=1.5cm]
  {$\mathrm{BM}_3 = |y_3 - r_{33}x_3|^2$};

\node[metric, above of=x2a, yshift=1cm]
  {$\mathrm{BM}_2 = |y_2 - r_{22}x_2 - r_{23}x_3|^2$};

\node[metric, above of=x1a, yshift=0.5cm]
  {$\mathrm{BM}_1 = |y_1 - r_{11}x_1 - r_{12}x_2 - r_{13}x_3|^2$};



\node[metric, above of=ml, xshift=-0.5cm,yshift=1.4cm]
  {Select path with minimum $\mathrm{PM}_1$};

\node[font=\footnotesize,above of=x3a, yshift=2cm] {Stage 1: $x_3$};
\node[font=\footnotesize,above of=x2a, yshift=2cm] {Stage 2: $x_2$};
\node[font=\footnotesize,above of=x1a, yshift=2cm] {Stage 3: $x_1$};

\end{tikzpicture}
\caption{Spatial Trellis Interpretation of 3x3 MIMO (64QAM). Arc represents select min ({\small survivor}) out of $|\mathcal{X}|$ outgoing paths}
\label{fig:placeholder}
\end{figure}

In fig. (1), in stage 1, there are $|\mathcal{X}|$ nodes and $BM_3$ is computed for each. At Stage 2, for each $x_3$ hypothesis, $x_2$ for all possible $|\mathcal{X}|$ candidate values are evaluated. The $x_2$ minimizing $\mathrm{BM}_2(x_2,x_3)$ is retained per $x_3$, resulting in $|\mathcal{X}|$ surviving $(x_3,x_2)$ hypotheses. This is mathematically equivalent to finding $x_2$ using equation 11 for the given $x_3$. This conditional minimization does not collapse the trellis and preserves multiple hypothesis for subsequent stages. $PM$ metric is also computed for the $|\mathcal{X}| (x_3,x_2)$ pairs. And this is repeated till the final stage. This is where the multiple hypothesis and trellis (MHT) part of MT-MHT-MD comes from.

\begin{figure}
\centering
\begin{tikzpicture}[node distance=1.8cm, every node/.style={circle, draw, minimum size=6mm}]
\node (r) {Root};

\node[right of=r] (x3) {$x_3$};
\node[right of=x3] (x2) {$x_2$};
\node[right of=x2] (x1) {$x_1$};

\draw[->] (r) -- node[above] {64} (x3);
\draw[->] (x3) -- node[above] {64} (x2);
\draw[->] (x2) -- node[above] {64} (x1);
\end{tikzpicture}
\caption{Multi-Pivot in 3x3 MIMO 64QAM: Search one pivoting on $x_3$ }
\label{fig:placeholder1}

\begin{tikzpicture}[node distance=1.8cm, every node/.style={circle, draw, minimum size=6mm}]
\node (r) {Root};

\node[right of=r] (x2) {$x_2$};
\node[right of=x2] (x1) {$x_1$};
\node[right of=x1] (x3) {$x_3$};

\draw[->] (r) -- node[above] {64} (x2);
\draw[->] (x2) -- node[above] {64} (x1);
\draw[->] (x1) -- node[above] {64} (x3);

\end{tikzpicture}
\caption{Multi-Pivot in 3x3 MIMO 64QAM: Search two pivoting on $x_2$ }
\label{fig:placeholder2}

\begin{tikzpicture}[node distance=1.8cm, every node/.style={circle, draw, minimum size=6mm}]
\node (r) {Root};

\node[right of=r] (x1) {$x_1$};
\node[right of=x1] (x3) {$x_3$};
\node[right of=x3] (x2) {$x_2$};

\draw[->] (r) -- node[above] {64} (x1);
\draw[->] (x1) -- node[above] {64} (x3);
\draw[->] (x3) -- node[above] {64} (x2);

\end{tikzpicture}
\caption{Multi-Pivot in 3x3 MIMO 64QAM: Search three pivoting on $x_1$ }
\label{fig:placeholder3}
\end{figure}

The above figure further illustrates the proposed interpretation of ML MIMO detection as a spatial trellis traversal problem. Each decoding order induces a distinct trellis, where all constellation hypotheses are retained at each depth. Crucially, the correct ML path does not dominate until full depth due to residual inter-layer interference, motivating delayed pruning.

ML MIMO detection can be viewed as sequence estimation over a spatial ISI channel.
Different column permutations of the channel matrix induce different spatial trellises. 
While all trellises correspond to the same ML objective, their intermediate metrics
evolve differently. Retaining all constellation hypotheses until sufficient trellis
depth ensures that the true ML path is not prematurely eliminated. This is where the Multi-Pivot (MP) part of MP-MHT-MD comes from. By virtue of this multi-pivot construction, competitor bit is always assured for soft llr computation.

The proposed algorithm may be interpreted as reverse-viterbi in the sense the pruning happens on the possible outgoing paths for each of the nodes at any stage $k$, $k = N_t-1, N_t-2,...,1$.

For $2 \times 2$, the 2 pivots with full hypotheses are as follows. It can be inferred that the $2|\mathcal{X}|$ list obtained this way actually contains the ML and the closest competitor pair for all bits in terms of euclidean distance as in the ML solution of size $|\mathcal{X}|^2$ and hence the soft/hard decisions are ML optimal (in max log sense) [1][2]. 

For $N_t \times N_t$, the solution is very close to hard ML and the soft llrs are also guaranteed for all bits because of the multi-pivot, multi-hypothesis approach. But they are over-estimated because the closest competitor bit/symbol pair is not guaranteed to be in the $N_t |\mathcal{X}|$ list. However forward error correction algorithms such as ldpc/turbo may tolerate the over-estimated llrs.

Figure 1 may also be interpreted as a classical ML sequence estimation (MLSE) problem over an equivalent spatial ISI channel with slight modification. In this interpretation, all candidate symbols at Stage 1 are considered and the branch metric $BM_3$ is computed. For each node at Stage 2, the best incoming path is then retained by minimizing the accumulated metric $BM_2+BM_3$, analogous to conventional Viterbi decoding. The same procedure is repeated at Stage 3 to obtain the final ML solution. However, this formulation requires evaluating $|\mathcal{X}|^2$ metrics per stage, resulting in a complexity that scales quadratically with the constellation size for any MIMO configuration. When extended to a multi-pivot formulation, the overall complexity increases to $N_t |\mathcal{X}|^2$, which remains significantly higher than that of the proposed detector. In contrast, the proposed algorithm exploits the upper-triangular structure obtained from QR decomposition, together with Viterbi-like pruning, to achieve a substantially lower and deterministic complexity.

\subsection{Algorithm}
The proposed algorithm is illustrated in Algorithm 1 section below. Please note that step 11 in the algorithm is effectively same as solving for $x_k$ using the equation corresponding to that layer after doing QR application (as illustrated by step 12).

\begin{algorithm}
\caption{Multi-Pivot Multiple-Hypothesis Trellis-Based MIMO Detection}
\label{alg:mpmh}
\begin{algorithmic}[1]
\REQUIRE Received vector $\mathbf{y}$, channel matrix $\mathbf{H}$, constellation $\mathcal{X}$, noise variance $\sigma^2$
\ENSURE Hard ML estimate $\hat{\mathbf{x}}$, soft LLRs

\STATE Initialize candidate list $\mathcal{L} \gets \emptyset$

\FOR{each pivot layer $p \in \{N, N-1, \ldots, 1\}$}

    \STATE Reorder columns of $\mathbf{H}$ such that symbol $x_p$ is the \textbf{first enumerated layer}
    \STATE Perform QR decomposition to obtain $\mathbf{R}$ and transformed $\mathbf{y}$

    \COMMENT{Stage 1: enumerate pivot-layer hypotheses}
    \FOR{each $x_p \in \mathcal{X}$}
        \STATE Compute branch metric:
        \[
        \text{BM}_p(x_p) = |y_p - r_{pp} x_p|^2
        \]
        \STATE Initialize path $(x_p)$ with $\text{PM}_p = \text{BM}_p(x_p)$
    \ENDFOR

    \COMMENT{Stages 2 to $p$: conditional best-symbol selection}
    \FOR{$k = p-1$ down to $1$}
        \FOR{each retained hypothesis path}
            \STATE For all $x_k \in \mathcal{X}$, compute the conditional branch metric and select:
            \[
            \text{BM}_k(x_k) = \left| y_k - \sum_{j=k}^{p} r_{kj} x_j \right|^2, \hat{x}_k = \arg\min_{x_k \in \mathcal{X}} \text{BM}_k(x_k)
            \]
            \STATE OR equivalently compute (instead of step 11):
            \[ 
            \hat{x}_k = slice( (y_k - \sum_{j=k+1}^{p} r_{kj} x_j)/r_{kk} )
            \]
            (only the selected symbol $\hat{x}_k$ is retained; no alternative hypotheses are expanded at this stage.)
            \STATE Update path metric:
            \[
            \text{PM}_k = \text{PM}_{k+1} + \text{BM}_k(\hat{x}_k)
            \]
        \ENDFOR
    \ENDFOR
    \STATE Store all completed paths $(\mathbf{x}, \text{PM}_1)$ in $\mathcal{L}$

\ENDFOR

\COMMENT{Final ML decision}
\STATE Select:
\[
\hat{\mathbf{x}} = \arg\min_{\mathbf{x} \in \mathcal{L}} \|\mathbf{y} - \mathbf{R}\mathbf{x}\|^2
\]

\COMMENT{Soft-output computation}
\FOR{each layer $k$ and bit position $b$}
    \STATE
    \[
    \text{LLR}_{k,b} =
    \min_{\mathbf{x}: b_{k}=1} \|\mathbf{y} - \mathbf{R}\mathbf{x}\|^2
    -
    \min_{\mathbf{x}: b_{k}=0} \|\mathbf{y} - \mathbf{R}\mathbf{x}\|^2
    \]
\ENDFOR

\RETURN $\hat{\mathbf{x}}$, LLRs
\end{algorithmic}
\end{algorithm}

\section{Results}

Simulations were run in a simplified framework with independent and identically distributed (iid) rayleigh channel with QPSK, 16-QAM, 64-QAM and 256-QAM for $2 \times 2$, $3 \times 3$, $4 \times 4$, $6 \times 6$ and $8 \times 8$ without any error control coding. Brute-force ML, proposed method with different list sizes and QR-ZF are simulated and BER vs SNR is plotted for each method.

\begin{figure}[!htbp]
    \includegraphics[width=1\linewidth]{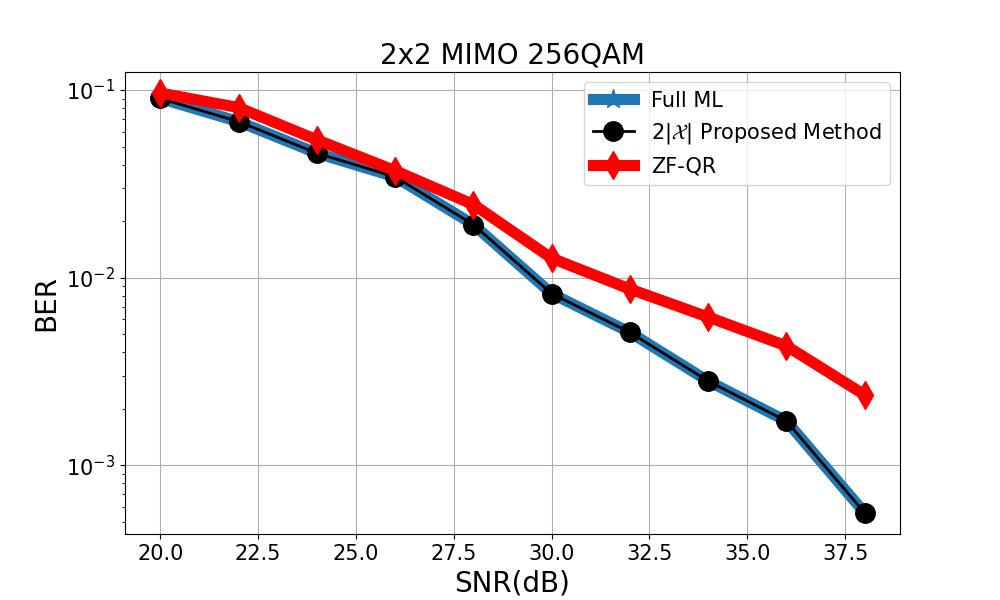}
    \caption{BER performance for 2×2 MIMO 256QAM: Full ML vs Reduced-Search ($2 |\mathcal{X}|$) vs ZF-QR}
    \label{fig:placeholder}
\end{figure}

\begin{figure}[!htbp]
    \centering
    \includegraphics[width=1\linewidth]{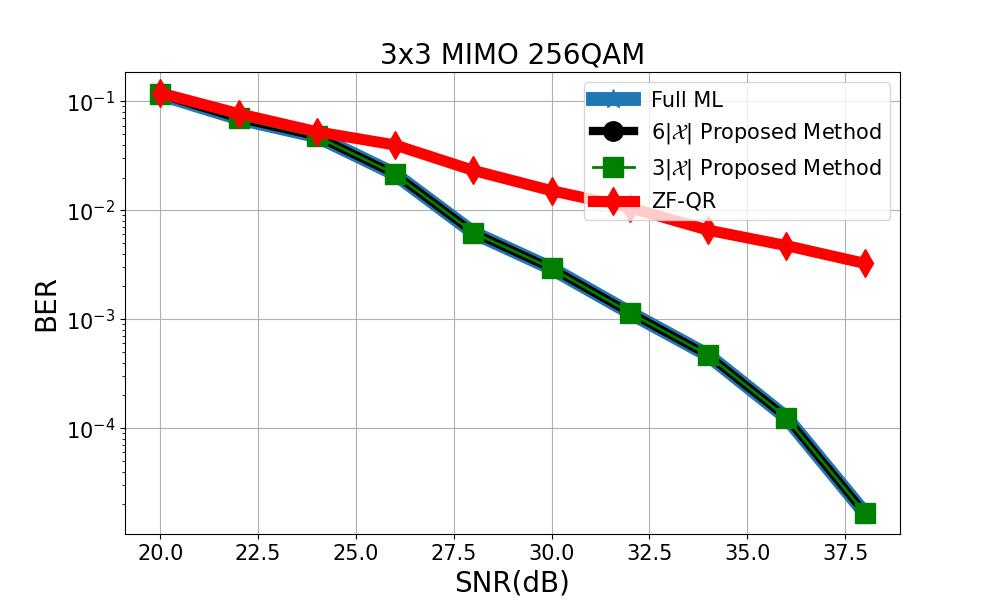}
    \caption{BER performance for 3x3 MIMO 256QAM: Full ML vs Reduced-Search ($6 |\mathcal{X}|$ and $3 |\mathcal{X}|$) vs ZF-QR}
    \label{fig:placeholder}
\end{figure}


\begin{figure}[!htbp]
    \centering
    \includegraphics[width=1\linewidth]{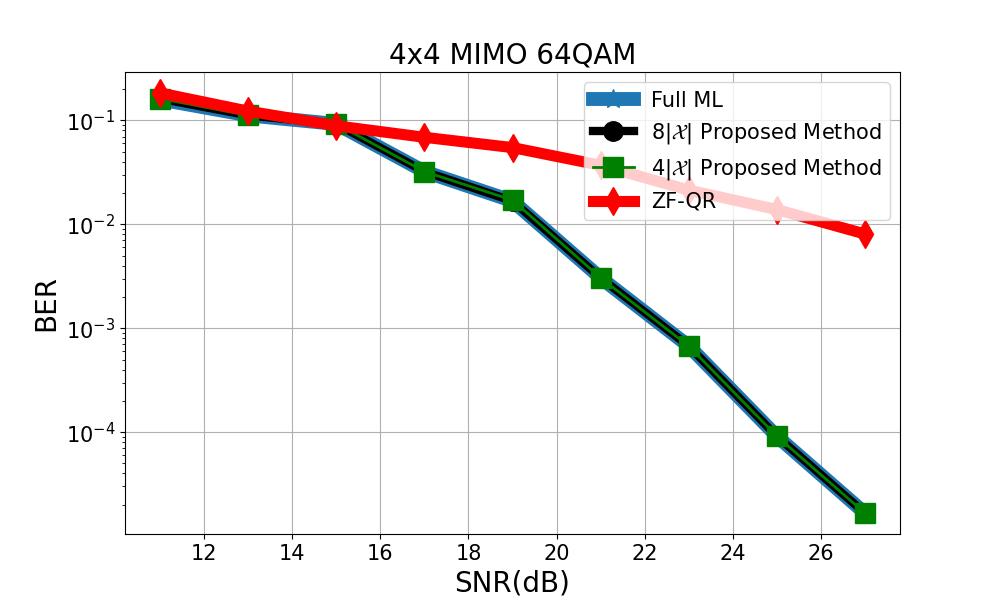}
    \caption{BER performance for 4×4 MIMO 64-QAM: Full ML vs Reduced-Search ($8 |\mathcal{X}|$ and $4 |\mathcal{X}|$) vs ZF-QR }
    \label{fig:placeholder}
\end{figure}

\begin{figure}[!htbp]
    \centering
    \includegraphics[width=1\linewidth]{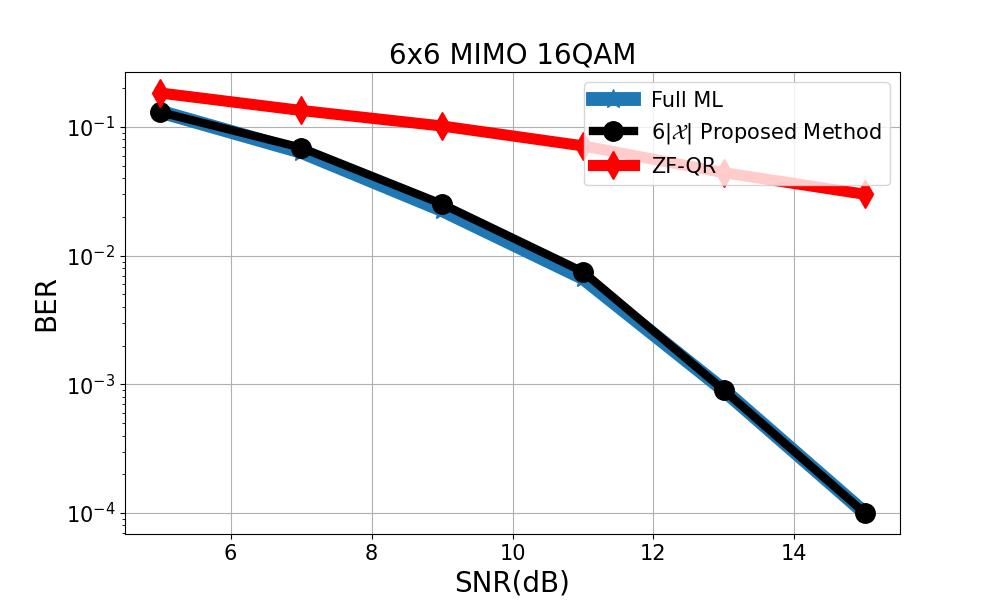}
    \caption{BER performance for 6×6 MIMO 16-QAM: Full ML vs Reduced-Search ($6 |\mathcal{X}|$)}
    \label{fig:placeholder}
\end{figure}

\begin{figure}[!htbp]
    \centering
    \includegraphics[width=1\linewidth]{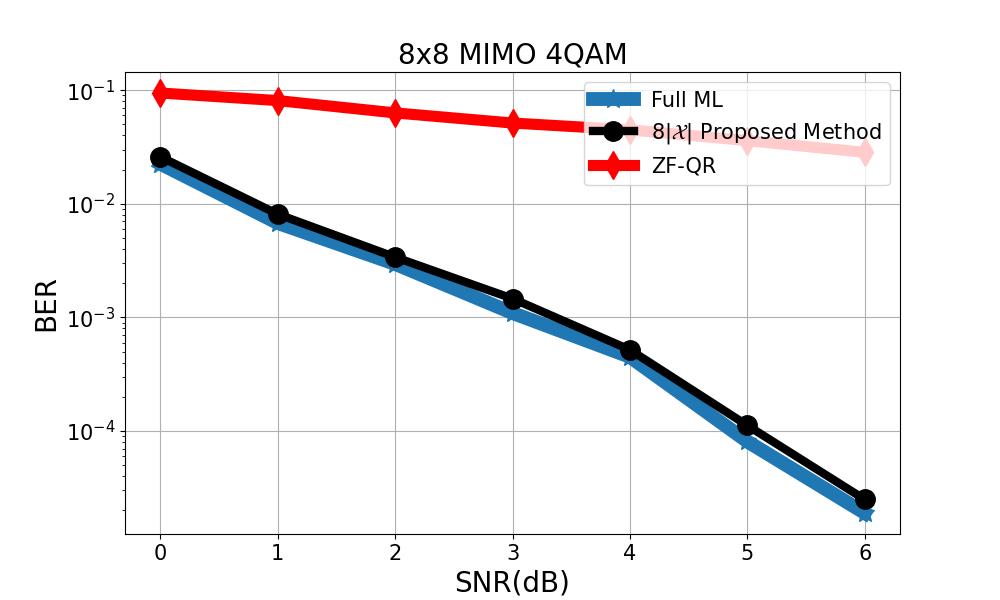}
    \caption{BER performance for 8×8 MIMO QPSK: Full ML vs Reduced-Search ($8 |\mathcal{X}|$)}
    \label{fig:placeholder}
\end{figure}

\begin{figure}[!htbp]
    \centering
    \includegraphics[width=1\linewidth]{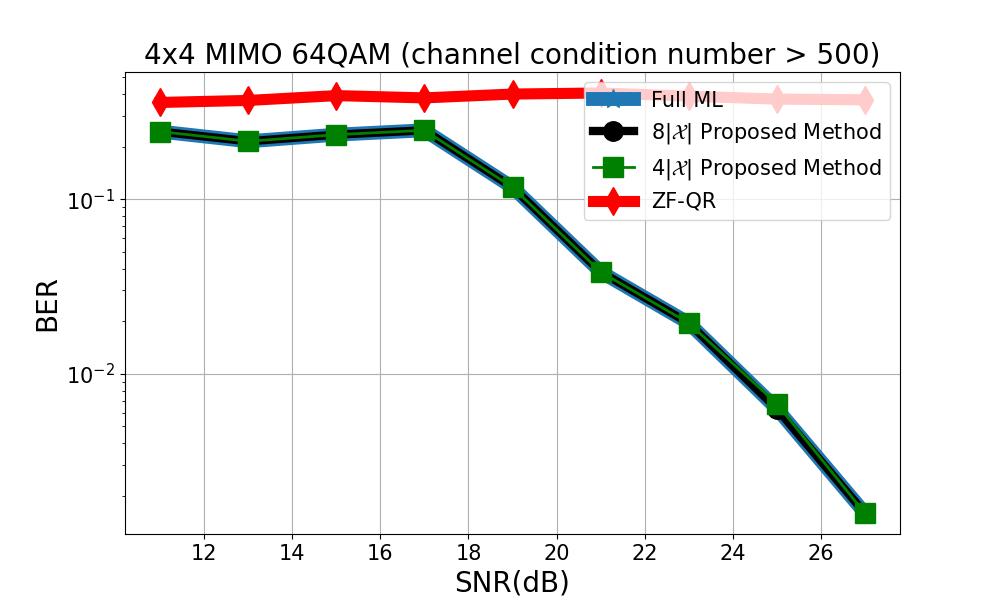}
    \caption{BER performance for 4×4 MIMO 64QAM with channel condition number $>$ 500: Full ML vs Reduced-Search ($8 |\mathcal{X}|$ and $4 |\mathcal{X}|$) vs ZF-QR}
    \label{fig:placeholder}
\end{figure}

\section{Key Observations and Insights}
\begin{itemize}
\item For $2 \times 2$ MIMO, the proposed method is exactly same as ML in hard decision and soft-llrs in max log sense but with list size of only $2 |\mathcal{X}|$ corresponding to ordering of $(x_1,x_2)$ and $(x_2,x_1)$.

\item For $3 \times 3$ MIMO, the proposed method can be applied with $3! = 6$ times $|\mathcal{X}|$ list size or with $3 |\mathcal{X}|$ list size. The former corresponds to ordering of $(x_1,x_2,x_3)$, $(x_1,x_3,x_2)$, $(x_2,x_3,x_1)$, $(x_2,x_1,x_3)$, $(x_3,x_1,x_2)$ and $(x_3,x_2,x_1)$. The latter corresponds to some further simplifications to reduce complexity - the ordering being $(x_1,x_2,x_3)$, $(x_2,x_3,x_1)$ and $(x_3,x_1,x_2)$.

\item For upto $4 \times 4$ MIMO, there is no noticeable degradation in BER performance for the proposed methods compared to ML. For $6 \times 6$ MIMO and above, there is slight (still less than 0.1-0.2dB) degradation only.

\item For $N_t \times N_t$ MIMO, the proposed method requires $N_t$ pivots and can be applied with $N_t!  |\mathcal{X}|$ list size (full hypothesis for all layers but with all combinations of other layers) or with much lesser complexity as $N_t |\mathcal{X}|$ list size (full hypothesis for each of the $N_t$ layers)

\item Reduced-search detection recovers ML-equivalent hard decisions across a wide range of SNRs, even under moderately ill-conditioned channel realizations.

\item Dominant error events are driven by single-layer symbol deviations (though not shown here); multi-layer error events occur with significantly lower probability and contribute marginally to overall BER.

\item High channel condition numbers increase noise enhancement but do not necessarily increase the dimensionality of dominant ML error events.

\item The effective ML candidate set in practical fading channels grows approximately linearly with constellation size and number of transmit layers, despite exponential worst-case complexity.

\item Competitors for calculating soft-output LLRs is always guaranteed because of the multi-pivot approach. For $ > 2 \times 2$ configuration, soft-output LLRs may exhibit overconfidence due to reduced hypothesis enumeration especially in channels with high condition number; however, these errors are structured and can be mitigated via condition-number-aware scaling or LLR clipping.

\item When combined with LDPC or Turbo decoding, the decoders may be able to tolerate moderate LLR overestimation and hence the proposed method can still achieve near-ML coded performance.

\item If interference is present in the system, one can apply Interference Rejection Combining (IRC) followed by the proposed method resulting in reduced complexity IRC-ML. In addition, general pre-processing steps like lattice reduction can be performed before the proposed method is applied.

\item Sphere decoding aggressively prunes early layers and risks discarding dominant ML competitors, whereas the proposed method retains full constellation hypotheses per layer, preserving ML-relevant candidates.

\item The proposed detector exhibits deterministic and fixed complexity, in contrast to the highly variable and tail-dominated complexity of sphere decoding, enabling hardware-friendly real-time implementation, particularly on parallel architectures such as GPUs.

\end{itemize}

The trade-offs of the various detection algorithms are presented in Table 1.

\begin{table}[t]
\fontsize{5.5pt}{5.5pt}\selectfont
\centering
\caption{Comparison of MIMO Detection Algorithms}
\label{tab:comparison}
\begin{tabular}{|l|c|c|c|}
\hline
\textbf{Comparison} 
& \textbf{ML} 
& \textbf{Proposed Method} 
& \textbf{Sphere Decoding} \\ \hline

Performance 
& Optimal 
& Near-optimal 
& Near-optimal \\ \hline

Complexity 
& High (exponential) 
& Medium (linear in $|\mathcal{X}|$) 
& Variable (polynomial on average) \\ \hline

Time Bound 
& Not practical 
& Deterministic
& Not time-bounded \\ \hline

Soft-LLRs 
& Yes 
& Yes 
& Not guaranteed \\ \hline

\end{tabular}
\end{table}

\section{Looking back}
Although the QR-decomposed MIMO detection problem has long been recognized as algebraically equivalent to an ISI channel, this equivalence has typically been interpreted in a strict trellis sense, leading to the conclusion that Viterbi-style decoding would require a state space of size $|\mathcal{X}|^{N_t-1}$ and is therefore impractical. As a result, prior work has largely focused on sphere decoding and K-best algorithms that rely on early pruning and variable-complexity search.

In contrast, the present work revisits this equivalence by exploiting the causal, upper-triangular structure induced by QR decomposition. This structure ensures that, for a fixed hypothesis at a given layer, the optimal predecessor at the next layer is uniquely determined by metric minimization, thereby eliminating the need to retain multiple competing paths per node. This subtle but important distinction breaks the conventional ISI-trellis complexity barrier and enables a deterministic, reduced-search procedure that retains ML-like behavior while avoiding exponential growth in candidate paths. The resulting interpretation bridges classical MLSE intuition with practical MIMO detection and reveals complexity reductions that have not been explicitly identified in prior literature.

\section{Soft-LLR Reliability and Rank-Aware Scaling}

While the proposed reduced-complexity detector achieves near-ML hard decision performance, the soft log-likelihood ratios (LLRs) obtained from a reduced candidate list may be over-estimated in certain scenarios, particularly when the closest competing hypothesis for a given bit lies far from the minimum-distance candidate in the reduced list. This behavior is inherent to any reduced-search detector and is not specific to the proposed approach.

To address this, a rank-aware LLR scaling mechanism is introduced that exploits the intrinsic geometry of the reduced search space without requiring any additional channel-dependent heuristics.

\subsection{Soft LLR Computation from Reduced Candidate List}

Let $\mathcal{C} = \{ \mathbf{x}_k \}_{k=1}^{N_c}$ denote the reduced candidate list generated by the proposed detector, with corresponding Euclidean distance metrics $\{ d_k \}_{k=1}^{N_c}$, sorted such that
\begin{equation}
d_{(1)} \le d_{(2)} \le \cdots \le d_{(N_c)} .
\end{equation}

For stream $t$ and bit position $b$, the soft LLR is computed as
\begin{equation}
\text{LLR}_{t,b} = d_{1} - d_{0},
\end{equation}
where
\begin{align}
d_0 &= \min_{k : b_{t,b}(\mathbf{x}_k)=0} d_k, \\
d_1 &= \min_{k : b_{t,b}(\mathbf{x}_k)=1} d_k.
\end{align}

\subsection{Rank-Based Confidence Measure}

Let $r_{t,b}$ denote the rank position (in the sorted metric list) of the closest competing hypothesis for bit $(t,b)$, i.e., the candidate achieving $\min(d_0,d_1)$ corresponding to the opposite bit value of the ML decision. A normalized rank measure is defined as
\begin{equation}
\alpha_{t,b} = \frac{r_{t,b}-1}{N_c-1}, \qquad \alpha_{t,b} \in [0,1].
\end{equation}

A small value of $\alpha_{t,b}$ indicates that the closest competing hypothesis is near the ML candidate and the corresponding LLR is reliable, whereas larger values indicate weaker competition and potential over-confidence.

\subsection{Rank-Aware LLR Scaling}

The final scaled LLR is obtained as
\begin{equation}
\text{LLR}_{t,b}^{\text{scaled}} =
\text{LLR}_{t,b} \cdot g(\alpha_{t,b}),
\end{equation}
where $g(\cdot)$ is a monotonic decreasing scaling function.

Two practical choices are considered:

\subsubsection{Linear Scaling}
\begin{equation}
g(\alpha) = 1 - \beta \alpha,
\end{equation}
where $\beta \in [0,1]$ controls the aggressiveness of the scaling.

\subsubsection{Exponential Scaling}
\begin{equation}
g(\alpha) = \exp(-\gamma \alpha),
\end{equation}
where $\gamma > 0$ determines the decay rate.

Both forms preserve LLR magnitude when the competing hypothesis is close to the ML candidate and progressively reduce over-confidence when the competitor appears deeper in the reduced list.

\subsection{Discussion}

The proposed rank-aware scaling leverages information already available from the reduced-search detector and introduces negligible additional complexity. Unlike channel-condition-number-based scaling methods, the proposed approach adapts on a per-bit basis and directly reflects the geometry of the reduced ML search space. Simulation results indicate that this scaling improves soft-decoding robustness when combined with LDPC or turbo decoders, while preserving near-ML hard decision performance.

\section{Practical Deployment Scenarios and Robustness Analysis}

The MP-MHT decoder is particularly relevant for modern cellular architectures where linear receivers hit fundamental performance barriers. We highlight two critical industrial use cases where the multi-hypothesis search provides a non-linear gain.

\subsection{Full-Rank MU-MIMO in FDD 4T4R Systems}
Traditional mobile traffic has been historically asymmetric, favoring the downlink. However, the emergence of AI and immersive technologies—such as Generative AI agents and edge-based AR/VR—has shifted the demand toward uplink-heavy, latency-sensitive traffic patterns. 

In current sub-3GHz FDD deployments ($N_r=4$), achieving the spectral efficiency required for these services necessitates scheduling four separate UEs in a single resource block ($N_t=4$). In this $N_r = N_t$ configuration, the receiver has zero additional Degrees-of-Freedom (DoF) for interference nulling. Linear receivers like Zero-Forcing (ZF) are highly sensitive to CSI aging and noise enhancement, often failing to support the high-order constellations (e.g., 1024-QAM, 256-QAM) required for high-fidelity AR/VR streams. The MP-MHT decoder overcomes this by resolving the four users' symbols jointly through the trellis structure, preserving LLR fidelity where linear methods suffer significant performance degradation.

\subsection{8-Layer Massive MIMO and the "Keyhole" Challenge}
In 5G-Advanced and 6G Massive MIMO ($N_r \ge 32$), achieving an 8-layer multiplexing gain is a primary goal for high-capacity urban cells. However, diffraction through narrow apertures in street canyons can create Keyhole (Pin-hole) channels. In such scenarios, even physically separated UEs share a common spatial path, resulting in a near-singular channel matrix where correlation $\rho \to 1.0$. 

Under these conditions, the MP-MHT decoder avoids explicit matrix inversion and treats it as a path-metric optimization problem. By exploring the trellis for the symbol sequence that minimizes the global Euclidean distance, it can effectively distinguish between layers that appear spatially identical to a linear filter.

\section{Further Work}
Several directions for future research remain open. Lattice-reduction-based
preprocessing may be combined with the proposed detection framework to further
improve robustness in highly spatially correlated or ill-conditioned channels.
The current study focuses on single-user MIMO scenarios with identical constellations
across all spatial layers. An important extension is the evaluation of the proposed
algorithm in multi-user MIMO (MU-MIMO) and heterogeneous SU-MIMO settings, where
different users or layers may employ distinct modulation and coding schemes. Some optimizations/list size reductions when precoding is employed (or in general) can also be explored. Condition number based switching between ZF/MMSE and the proposed method is another area. Finally, the broader applicability of the proposed multi-pivot, multiple-hypothesis
search framework to lattice problems beyond MIMO detection—such as the Closest
Vector Problem (CVP)—is a promising direction for further exploration.

\section{Conclusions}

\textit{The proposed method effectively bridges the performance gap with respect to ML but at significantly lower complexity, even though exact ML remains computationally intractable.}

This work demonstrates that:
\begin{itemize}
\item Exponential ML complexity is unnecessary in practical MIMO systems
\item Near-ML hard and soft decisions are achievable with linear list sizes
\item The approach scales cleanly from $2\times2$ to $8\times8$ MIMO
\end{itemize}

The author hopes that the insights and interpretations presented in this work will be of interest to both industry and academia, and may stimulate further investigation into related detection, inference, and lattice-search problems.


\begin{thebibliography}{1}
\bibliographystyle{IEEEtran}

\bibitem{ref1}
Logeshwaran Vijayan, Partha Sarathy Murali, Sundaram Vanka,  "Reduced complexity maximum likelihood decoder for MIMO communications", Patent US7965782B1, June 21, 2011.

\bibitem{ref7}
M. Siti and M. P. Fitz, "A Novel Soft-Output Layered Orthogonal Lattice Detector for Multiple Antenna Communications," 2006 IEEE International Conference on Communications, Istanbul, Turkey, 2006, pp. 1686-1691, doi: 10.1109/ICC.2006.254962.

\end{thebibliography}
\end{document}